\documentclass[aps,prd,amsmath,amssymb,superscriptaddress,preprintnumbers,twocolumn,10pt,nofootinbib]{revtex4-1}
\usepackage{graphicx}
\usepackage{dcolumn}
\usepackage{bm}
\usepackage{amssymb}
\usepackage{latexsym}
\usepackage{booktabs}
\usepackage{amsmath}
\usepackage{multirow}
\usepackage{url}
\usepackage{footnote}
\usepackage{float}
\usepackage{threeparttable}
\usepackage[colorlinks=true, linkcolor=blue, citecolor=blue]{hyperref}
\usepackage[bottom]{footmisc}

\usepackage[normalem]{ulem}
\usepackage{color}
\usepackage{array}
\usepackage{enumerate}
\usepackage{adjustbox}

\usepackage{makecell}
\usepackage{diagbox}
\usepackage{epstopdf}
\usepackage{epsfig}
\usepackage{longtable}
\usepackage{supertabular}
\usepackage{algorithm}
\usepackage{pifont}
\usepackage{algorithmic}
\usepackage{changepage}
\usepackage{setspace}
\IfFileExists{orcidlink.sty}{\usepackage{orcidlink}}{%
  \newcommand{\orcidlink}[1]{%
    \href{https://orcid.org/##1}{%
      \includegraphics[height=1.6ex]{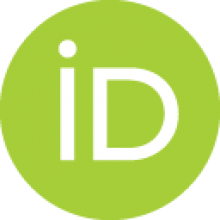}%
    }%
  }%
}

\begin{document}

\title{Model-independent late-universe measurements of $H_0$ and $\Omega_\mathrm{K}$ with the parametrization based on cosmic age-improved inverse distance ladder}

\author{Guo-Hong Du\orcidlink{0009-0005-6921-3201}}
\affiliation{Liaoning Key Laboratory of Cosmology and Astrophysics, College of Sciences, Northeastern University, Shenyang 110819, China}

\author{Tian-Nuo Li\orcidlink{0009-0004-6982-4021}}
\affiliation{Liaoning Key Laboratory of Cosmology and Astrophysics, College of Sciences, Northeastern University, Shenyang 110819, China}

\author{Jia-Le Ling\orcidlink{0009-0000-8052-0994}}
\affiliation{Institute of Theoretical Physics, Chinese Academy of Sciences, Beijing 100190, China}

\author{Yan-Hong Yao\orcidlink{0000-0001-5283-3635}}
\affiliation{Institute of Fundamental Physics and Quantum Technology, Department of Physics, School of Physical Science and Technology, Ningbo University, Ningbo 315211, China}

\author{Jing-Fei Zhang\orcidlink{0000-0002-3512-2804}}
\affiliation{Liaoning Key Laboratory of Cosmology and Astrophysics, College of Sciences, Northeastern University, Shenyang 110819, China}

\author{Xin Zhang\orcidlink{0000-0002-6029-1933}}\thanks{Corresponding author}\email{zhangxin@neu.edu.cn}
\affiliation{Liaoning Key Laboratory of Cosmology and Astrophysics, College of Sciences, Northeastern University, Shenyang 110819, China}
\affiliation{MOE Key Laboratory of Data Analytics and Optimization for Smart Industry, Northeastern University, Shenyang 110819, China}
\affiliation{National Frontiers Science Center for Industrial Intelligence and Systems Optimization, Northeastern University, Shenyang 110819, China}

\begin{abstract}

The standard $\Lambda{\rm CDM}$ model has encountered serious challenges and the $H_0$ tension has become more significant with increasingly precise cosmological observations. Meanwhile, inconsistencies in measurements of the curvature parameter $\Omega_\mathrm{K}$ between different datasets also have emerged. In this work, we employ two global and cosmic age-based parameterizations, PAge and MAPAge, to perform model-independent measurements of the Hubble constant $H_0$ and $\Omega_\mathrm{K}$ by utilizing the inverse distance ladder (IDL). To construct the PAge-improved IDL, we utilize the strong gravitational lensing (SGL), cosmic chronometers (CC), and gamma ray bursts (GRB) data to calibrate the latest DESI DR2 baryon acoustic oscillation and DESY5 or DES-Dovekie type Ia supernova data. Our analysis indicates that DESI+DES-Dovekie+SGL+CC+GRB gives $H_0=72.20\pm 1.00\,{\rm km}~{\rm s}^{-1}~{\rm Mpc}^{-1}$ in the MAPAge model, reducing the $H_0$ tension to the $0.6\sigma$ level. Extending to the MAPAge$+\Omega_{\rm K}$ model, we obtain $\Omega_\mathrm{K}=0.005\pm 0.037$, which suggests that current late-time data are consistent with a flat universe. Finally, the Bayesian analysis indicates that the present late-universe data provide weak to moderate evidence in favor of PAge and MAPAge relative to $\Lambda{\rm CDM}$.

\end{abstract}

\maketitle

\section{Introduction}

The Hubble constant $H_0$ and the curvature parameter $\Omega_\mathrm{K}$ are important and fundamental parameters in cosmology, where $H_0$ represents the current expansion rate of the universe and $\Omega_\mathrm{K}$ determines the global spatial geometry of the universe (open, flat, or closed). However, the measurements of these two parameters currently remain in tension. The $H_0$ tension between early- and late-time universe is one of the most severe tensions in modern cosmology~\cite{Bernal:2016gxb,DiValentino:2021izs,Abdalla:2022yfr,CosmoVerseNetwork:2025alb}, reaching about the $4-6\sigma$ level. Specifically, local distance ladder measurements, such as SH0ES Cepheid-calibrated type Ia supernova (SN)~\cite{Riess:2021jrx}, tend to give $H_0 \sim 72-75\,\mathrm{km\,s^{-1}\,Mpc^{-1}}$. By contrast, early-universe data, particularly cosmic microwave background (CMB) observations from Planck~\cite{Planck:2018vyg} and the Atacama Cosmology Telescope (ACT)~\cite{ACT:2020gnv,ACT:2025llb}, yield values approximately $66-68\,\mathrm{km\,s^{-1}\,Mpc^{-1}}$ assuming the standard $\Lambda$-cold dark matter ($\Lambda$CDM) model. Furthermore, the measurements of $\Omega_\mathrm{K}$ also exhibit a discrepancy. The Planck CMB data alone yield $\Omega_\mathrm{K} = -0.044^{+0.018}_{-0.015}$, which deviates from zero at approximately the $3\sigma$ level~\cite{Planck:2018vyg}. However, the combination of CMB and baryon acoustic oscillation (BAO) data obtains $\Omega_\mathrm{K} = 0.0007\pm0.0019$, suggesting a spatially flat universe within the $1\sigma$ range~\cite{Planck:2018vyg}. This indicates an inconsistency between CMB and BAO data in measurements of $\Omega_\mathrm{K}$, commonly referred to as the curvature tension~\cite{Handley:2019tkm}.

The $H_0$ tension has prompted extensive debate. If it is not caused by systematic errors, it likely indicates new physics beyond the $\Lambda$CDM model~\cite{Zhang:2017aqn,Guo:2018ans,Planck:2019evm,Vagnozzi:2019ezj,DiValentino:2020vvd,Vagnozzi:2021tjv,Vagnozzi:2023nrq,Jin:2025dvf}. Various modifications have been proposed to resolve the $H_0$ tension, such as early dark energy, interacting dark energy, or modified gravity, and others~\cite{Boisseau:2000pr,Linder:2002et,Huang:2004wt,Farrar:2003uw,Zhang:2005kj,Zhang:2005hs,Zhang:2005yz,Cai:2004dk,Zhang:2005rg,Zhang:2005rj,Zhang:2006qu,Zhang:2004gc,Zhang:2007sh,Wang:2006qw,Ma:2007av,Zhang:2009un,DeFelice:2010aj,Capozziello:2011et,Fu:2011ab,Zhang:2012uu,Clifton:2011jh,Li:2014cee,Zhang:2015rha,Landim:2015hqa,Li:2015vla,Cai:2015emx,Wang:2016och,Costa:2016tpb,Chiang:2018xpn,Poulin:2018cxd,Feng:2017usu,Pan:2019gop,Pan:2020zza,Yao:2020hkw,Smith:2020rxx,Sekiguchi:2020teg,Yao:2020pji,Drepanou:2021jiv,Wang:2021kxc,Yin:2023srb,Yao:2022kub,Wang:2023mir,Li:2023gtu,Wang:2024vmw,Song:2022siz,Jiang:2025hco,Han:2024sxm,Mirpoorian:2024fka,Huang:2024erq,Zhang:2024rra,Song:2025ddm,Yin:2026gss}. Notably, most models assume a flat universe as the default prior. Relaxing this assumption affects distance inferences and can introduce degeneracies with $H_0$ or produce shifts in constraints. For example, a CMB-driven preference for negative curvature can exacerbate the $H_0$ tension~\cite{DiValentino:2020vvd,DiValentino:2021izs}. Therefore, given the persistent Hubble tension and the uncertainty about whether the universe is spatially flat, it is crucial to jointly measure $H_0$ and $\Omega_\mathrm{K}$.

The inverse distance ladder (IDL) provides a novel method for measuring $H_0$ and $\Omega_\mathrm{K}$ by calibrating the SN absolute magnitude $M_\mathrm{B}$ through anchoring the BAO sound horizon $r_\mathrm{d}$. To avoid early-time CMB priors that may be unhelpful for alleviating the $H_0$ tension, one can instead use late-time observations to calibrate BAO and SN. These include cosmic chronometer (CC) data~\cite{Cai:2021weh,Cai:2022dkh}, strong gravitational lensing (SGL)~\cite{Taubenberger:2019qna,Birrer:2020tax,Du:2023zsz,Zhang:2023eup}, gamma-ray bursts (GRB)~\cite{Demianski:2016zxi,Amati:2018tso}, and quasars~\cite{Lusso:2020pdb,Wei:2020suh}. However, when using the IDL, differences arising from specific cosmological model assumptions cannot be neglected.

To reduce dependence on cosmological models, many model-independent methods like Taylor expansions of the Hubble parameter $H(z)$~\cite{Zhang:2016urt,DES:2024ywx}, Padé approximations~\cite{Capozziello:2020ctn}, and Gaussian Processes~\cite{Jiang:2024xnu,Guo:2024pkx} have been developed. However, they suffer from high-redshift divergence~\cite{Cai:2022dkh}, unphysical poles~\cite{Capozziello:2020ctn}, or kernel sensitivity~\cite{OColgain:2021pyh,Kjerrgren:2021zuo}. In view of these limitations, \citet{Huang:2020mub} proposed a parameterization based on cosmic age (PAge), capturing the late-time expansion history with two external parameters while remaining a good approximation over a wide redshift range covered by $z\lesssim10^{4}$~\cite{Huang:2020mub,Luo:2020ufj}. By introducing a higher-order parameter $\eta_2$, the more accurate parameterization based on cosmic age (MAPAge) model further extends PAge to better characterize the expansion history~\cite{Huang:2021aku,Cai:2022dkh}. The PAge and MAPAge models can approximate cosmological distances and the expansion history to high precision~\cite{Huang:2020mub,Luo:2020ufj,Cai:2021weh,Cai:2022dkh}, making them effective tools for model-independent cosmological research.

Recently, the second data release (DR2) of the Dark Energy Spectroscopic Instrument (DESI) presented its latest three-year BAO measurements. The combination of DESI DR2 BAO, DESY5 SN, and CMB measurements from Planck and ACT favors dynamical dark energy at the $4.2\sigma$ level~\cite{DESI:2025zgx}. This significant deviation from $\Lambda$CDM has sparked extensive debates~\cite{Giare:2024smz,Li:2024qso,Sabogal:2024yha,Yang:2024kdo,RoyChoudhury:2024wri,Escamilla:2024ahl,Dinda:2024ktd,Li:2024qus,Huang:2025som,Wu:2025wyk,Li:2025ula,Li:2025ops,Ling:2025lmw,Pang:2025lvh,Ozulker:2025ehg,Cheng:2025lod,Silva:2025twg,Chen:2025wwn,Wu:2024faw,Du:2024pai,Jiang:2024viw,Ye:2024ywg,Feng:2025mlo,Wang:2025dtk,Cai:2025mas,Yang:2025ume,Li:2025eqh,RoyChoudhury:2025dhe,Li:2025htp,Giare:2025ath,Liu:2025myr,Yao:2025kuz,Cheng:2025cmb,Li:2025dwz,Braglia:2025gdo,Li:2025muv,Paul:2025wix,Yang:2025oax,Wang:2025vtw,Colgain:2025fct,Hogas:2025mii,Du:2025xes,Wang:2024dka,Li:2025owk,Pan:2025qwy,Lee:2025pzo,Du:2025iow,Li:2026hwq,Wang:2026sqy,Yao:2025twv,Sabogal:2025qhz,Barua:2025adv,Feng:2026pzs,Akita:2025txo,Millard:2026wnd,Wu:2026klh,Mukhopadhyay:2026fyk,Song:2025bio,Li:2025vuh,Li:2026xaz,Du:2026cly,Li:2026ldf,Du:2026qtq,Li:2026asg}. However, this quintom-type~\cite{Feng:2004ad} dynamical dark energy exacerbates the $H_0$ tension by predicting a lower $H_0$~\cite{DESI:2025zgx}. Moreover, the combination of DESI DR2 BAO and CMB data yields $\Omega_\mathrm{K}=0.023\pm0.011$ in the $\Lambda$CDM model, corresponding to a $2\sigma$ deviation from zero curvature. By contrast, within the dynamical dark energy framework the result instead points toward a spatially flat universe~\cite{DESI:2025zgx}. Consequently, in the context of the current DESI data, it is imperative to determine $H_0$ and $\Omega_\mathrm{K}$ in a model-independent scenario.

In this letter, we construct the IDL using DESI DR2 BAO, DESY5 and DES-Dovekie SN, TDCOSMO SGL, CC measurements, and a calibrated long-GRB sample. Employing the PAge and MAPAge models, we obtain model-independent measurements of $H_0$ and $\Omega_\mathrm{K}$, and perform Bayesian model comparisons. 

\section{Methodology}\label{sec2}

By expanding $Ht$ to the quadratic order in $t/t_0$, the PAge model is defined as~\cite{Huang:2020mub,Luo:2020ufj}
\begin{align}\label{eq:PAge}
Ht-\frac23=\left[p_{\mathrm{age}}-\frac23(1+\eta)\right]\frac{t}{t_0}+\frac23\eta\left(\frac{t}{t_0}\right)^2,
\end{align}
where $p_{\mathrm{age}} \equiv H_0t_0$ is a dimensionless parameter, and the parameter $\eta=1-\frac32p_\mathrm{age}^2(1+q_0)$ characterizes the deviation from the matter-dominated scenario ($Ht = 2/3$). By substituting $H(z) = -\mathrm{d}z/[(1+z)\mathrm{d}t]$ into Eq.~\eqref{eq:PAge}, we obtain the redshift-time relation $z(t)$ as
\begin{align}
1 + z = \left(\frac{p_{\mathrm{age}}}{H_0t}\right)^{\frac{2}{3}} \mathrm{e}^{\frac{1}{3}\left(1 - \frac{H_0t}{p_{\mathrm{age}}}\right)\left(3p_{\mathrm{age}} + \eta\frac{H_0t}{p_{\mathrm{age}}} - \eta - 2\right)}.
\end{align}
This model provides a sub-percent approximation for the late-time expansion history.

To achieve even higher accuracy and better flexibility across extended redshift ranges, the MAPAge model extends the expansion to the cubic order in cosmic time~\cite{Huang:2021aku,Cai:2022dkh}. The Hubble parameter in the MAPAge model is expressed as
\begin{align}\label{eq:MAPAge}
\frac{H}{H_0}=1+\frac23&\left[1-(\eta+\eta_2)\frac{H_0t}{p_\mathrm{age}}+\eta_2\left(\frac{H_0t}{p_\mathrm{age}}\right)^2\right]\nonumber\\
&\times\left(\frac{1}{H_0t}-\frac{1}{p_\mathrm{age}}\right),
\end{align}
where the additional parameter $\eta_2=1-\frac34p_\mathrm{age}^3(2+j_0+3q_0)$ is introduced to account for the present-day jerk parameter $j_0$. Similar to the PAge model, the evolution of $z(t)$ in the MAPAge framework is derived as
\begin{align}
1 + z = \left(\frac{p_{\mathrm{age}}}{H_0t}\right)^{\frac{2}{3}} 
& \exp\left\{\frac{1}{9}\left(1 - \frac{H_0t}{p_{\mathrm{age}}}\right) 
\left[9p_{\mathrm{age}} - 2\eta_2\left(\frac{H_0t}{p_{\mathrm{age}}}\right)^2 \right.\right. \nonumber\\ 
& \left.\left. + \left(3\eta+4\eta_2\right)\frac{H_0t}{p_{\mathrm{age}}} 
- \left(3\eta+2\eta_2+6\right)\right]\right\}.
\end{align}
Compared to PAge, MAPAge significantly reduces the relative error, making it more robust for distinguishing between models with subtle differences in expansion histories. For a comprehensive performance analysis, please refer to Ref.~\cite{Cai:2022dkh}.

We utilize the following late-time cosmological observations: the latest BAO measurements from DESI DR2 (\texttt{DESI})~\cite{DESI:2025zgx}, the SN data from the DESY5 sample (\texttt{DESY5})~\cite{DES:2024jxu} and its comprehensively re-calibrated sample (\texttt{DES-Dovekie})~\cite{DES:2025sig}, 32 CC measurements with total covariance matrix (\texttt{CC})~\cite{Jimenez:2003iv,Simon:2004tf,Stern:2009ep,Moresco:2012jh,Zhang:2012mp,Moresco:2015cya,Moresco:2016mzx,Ratsimbazafy:2017vga,Moresco:2020fbm,Vagnozzi:2020dfn,Borghi:2021rft,Moresco:2022phi}, seven well-studied SGL systems from the TDCOSMO collaboration (\texttt{SGL})~\cite{Suyu:2009by,Suyu:2013kha,H0LiCOW:2016qrm,H0LiCOW:2018tyj,H0LiCOW:2019xdh,Birrer:2020tax,H0LiCOW:2019mdu,DES:2019fny,Hogg:2023khs}, and 193 long GRBs standardized through the well-established Amati correlation (\texttt{GRB})~\cite{Amati:2018tso,Demianski:2019vzl}. For parameter inference and model comparison, we employ the \texttt{Cobaya} package~\cite{Torrado:2020dgo} alongside the \texttt{PolyChord} nested sampler~\cite{Handley:2015fda,Handley:2015vkr} to compute the Bayesian evidence, stopping when the remaining evidence drops below 0.1\% of the total. The posterior samples are then analyzed using \texttt{GetDist}~\cite{Lewis:2019xzd} to derive parameter constraints. For quantitative model comparison, we compute the Bayes factor in logarithmic space between two models, defined as $\ln\mathcal{B}_{ij}=\ln\mathcal{Z}_i-\ln\mathcal{Z}_j$, where $\ln\mathcal{Z}_i$ and $\ln\mathcal{Z}_j$ are the Bayesian evidences of models $i$ and $j$, respectively~\cite{Kass:1995loi}. We interpret the strength of evidence using Jeffreys scale~\cite{Trotta:2008qt}.

\begin{table*}[!htb]
\renewcommand\arraystretch{1.5}
\centering
\caption{The constraint results for the $\Lambda$CDM, PAge, MAPAge, $\Lambda\mathrm{CDM}+\Omega_{\rm K}$, PAge$+\Omega_{\rm K}$, and MAPAge$+\Omega_{\rm K}$ models obtained by the DESI, DESY5, DES-Dovekie, SGL, CC, and GRB data. Here, $H_{0}$ is in units of ${\rm km}~{\rm s}^{-1}~{\rm Mpc}^{-1}$ and $r_\mathrm{d}$ is in units of ${\rm Mpc}$.}
\label{table:St_new}
\resizebox{\textwidth}{!}{%
\setlength{\tabcolsep}{5pt}
\begin{tabular}{lccccccc}
\hline\hline
Model / Dataset & $H_0$ & $\eta\,\text{or}\,\Omega_\mathrm{m}$ & $\eta_2$ & $p_{\rm age}$ & $r_\mathrm{d}$ & $M_\mathrm{B}$ & $\Omega_\mathrm{K}$ \\
\hline

\multicolumn{8}{l}{$\boldsymbol{\Lambda\mathrm{CDM}}$} \\
DESI+DESY5+SGL             & $71.40^{+2.10}_{-1.80}$ & $0.3083\pm 0.0078$ & $\text{---}$ & $\text{---}$ & $141.2^{+3.5}_{-4.4}$ & $-19.288^{+0.065}_{-0.054}$ & $\text{---}$ \\
DESI+DESY5+CC              & $70.00^{+2.40}_{-2.10}$ & $0.3099^{+0.0067}_{-0.0079}$ & $\text{---}$ & $\text{---}$ & $143.9^{+4.1}_{-5.0}$ & $-19.330^{+0.075}_{-0.062}$ & $\text{---}$ \\
DESI+DESY5+GRB             & $74.50\pm 1.10$ & $0.3081\pm 0.0080$ & $\text{---}$ & $\text{---}$ & $135.3\pm 1.9$ & $-19.195\pm 0.031$ & $\text{---}$ \\
DESI+DESY5+SGL+CC          & $70.60^{+1.80}_{-1.20}$ & $0.3078\pm 0.0076$ & $\text{---}$ & $\text{---}$ & $142.9^{+2.5}_{-3.7}$ & $-19.312^{+0.055}_{-0.037}$ & $\text{---}$ \\
DESI+DESY5+SGL+CC+GRB      & $73.16\pm 0.89$ & $0.3069\pm 0.0075$ & $\text{---}$ & $\text{---}$ & $137.8\pm 1.6$ & $-19.234\pm 0.025$ & $\text{---}$ \\
DESI+DES-Dovekie+SGL+CC+GRB      & $73.34\pm 0.91$ & $0.3023\pm 0.0075$ & $\text{---}$ & $\text{---}$ & $137.9\pm 1.6$ & $-19.229\pm 0.026$ & $\text{---}$ \\
\hline

\multicolumn{8}{l}{$\boldsymbol{\mathrm{PAge}}$} \\
DESI+DESY5+SGL             & $70.10\pm 2.00$ & $0.2730\pm 0.0400$ & $\text{---}$ & $0.9518\pm 0.0075$ & $141.8\pm 4.1$ & $-19.304\pm 0.062$ & $\text{---}$ \\
DESI+DESY5+CC              & $68.40\pm 2.30$ & $0.2690\pm 0.0410$ & $\text{---}$ & $0.9502\pm 0.0070$ & $144.9^{+4.4}_{-5.5}$ & $-19.354^{+0.078}_{-0.071}$ & $\text{---}$ \\
DESI+DESY5+GRB             & $73.40\pm 1.30$ & $0.2760\pm 0.0420$ & $\text{---}$ & $0.9526\pm 0.0075$ & $135.2\pm 2.2$ & $-19.202\pm 0.035$ & $\text{---}$ \\
DESI+DESY5+SGL+CC          & $69.40^{+2.00}_{-1.70}$ & $0.2700\pm 0.0410$ & $\text{---}$ & $0.9515\pm 0.0070$ & $142.9^{+3.3}_{-4.0}$ & $-19.323^{+0.061}_{-0.050}$ & $\text{---}$ \\
DESI+DESY5+SGL+CC+GRB      & $72.16\pm 0.97$ & $0.2770\pm 0.0420$ & $\text{---}$ & $0.9531\pm 0.0073$ & $137.7\pm 1.6$ & $-19.240\pm 0.026$ & $\text{---}$ \\
DESI+DES-Dovekie+SGL+CC+GRB      & $72.63\pm 0.96$ & $0.3220\pm 0.0400$ & $\text{---}$ & $0.9589\pm 0.0068$ & $138.0\pm 1.6$ & $-19.235\pm 0.026$ & $\text{---}$ \\
\hline

\multicolumn{8}{l}{$\boldsymbol{\mathrm{MAPAge}}$} \\
DESI+DESY5+SGL             & $69.20^{+2.20}_{-1.90}$ & $0.0670^{+0.0840}_{-0.1100}$ & $0.54^{+0.25}_{-0.20}$ & $0.9361\pm 0.0089$ & $142.7^{+3.5}_{-4.7}$ & $-19.312^{+0.070}_{-0.054}$ & $\text{---}$ \\
DESI+DESY5+CC              & $68.30\pm 2.50$ & $0.0780^{+0.0900}_{-0.1000}$ & $0.50^{+0.23}_{-0.21}$ & $0.9362^{+0.0089}_{-0.0100}$ & $144.4^{+4.8}_{-5.8}$ & $-19.338^{+0.086}_{-0.073}$ & $\text{---}$ \\
DESI+DESY5+GRB             & $73.20\pm 1.30$ & $0.0910\pm 0.0940$ & $0.49^{+0.23}_{-0.21}$ & $0.9391\pm 0.0094$ & $135.1\pm 2.1$ & $-19.193\pm 0.035$ & $\text{---}$ \\
DESI+DESY5+SGL+CC          & $69.00^{+1.70}_{-1.40}$ & $0.0770\pm 0.0920$ & $0.52\pm 0.22$ & $0.9376\pm 0.0092$ & $143.2^{+2.7}_{-3.4}$ & $-19.318^{+0.052}_{-0.039}$ & $\text{---}$ \\
DESI+DESY5+SGL+CC+GRB      & $71.59\pm 0.94$ & $0.0930\pm 0.0910$ & $0.49\pm 0.21$ & $0.9391\pm 0.0090$ & $138.1\pm 1.6$ & $-19.240^{+0.026}_{-0.023}$ & $\text{---}$ \\
DESI+DES-Dovekie+SGL+CC+GRB      & $72.20\pm 1.00$ & $0.1830\pm 0.0950$ & $0.36\pm 0.22$ & $0.9488\pm 0.0094$ & $138.2\pm 1.7$ & $-19.234\pm 0.026$ & $\text{---}$ \\
\hline

\multicolumn{8}{l}{$\boldsymbol{\Lambda\mathrm{CDM}+\Omega_{\rm K}}$} \\
DESI+DESY5+SGL             & $69.50\pm 2.00$ & $0.2880\pm 0.0120$ & $\text{---}$ & $\text{---}$ & $143.9^{+3.7}_{-4.3}$ & $-19.333\pm 0.060$ & $0.087\pm 0.038$ \\
DESI+DESY5+CC              & $68.90\pm 2.30$ & $0.2930\pm 0.0110$ & $\text{---}$ & $\text{---}$ & $145.2^{+4.5}_{-5.2}$ & $-19.352^{+0.078}_{-0.070}$ & $0.071\pm 0.037$ \\
DESI+DESY5+GRB             & $74.80^{+1.20}_{-1.30}$ & $0.2880\pm 0.0110$ & $\text{---}$ & $\text{---}$ & $133.7\pm 2.1$ & $-19.173^{+0.032}_{-0.037}$ & $0.086\pm 0.036$ \\
DESI+DESY5+SGL+CC          & $69.20^{+1.60}_{-1.20}$ & $0.2870\pm 0.0110$ & $\text{---}$ & $\text{---}$ & $144.5^{+2.4}_{-3.2}$ & $-19.341^{+0.049}_{-0.037}$ & $0.090\pm 0.036$ \\
DESI+DESY5+SGL+CC+GRB      & $72.69\pm 0.93$ & $0.2960\pm 0.0100$ & $\text{---}$ & $\text{---}$ & $137.8\pm 1.6$ & $-19.239\pm 0.026$ & $0.054\pm 0.031$ \\
DESI+DES-Dovekie+SGL+CC+GRB      & $72.92\pm 0.91$ & $0.2941\pm 0.0098$ & $\text{---}$ & $\text{---}$ & $138.1\pm 1.6$ & $-19.235\pm 0.025$ & $0.041^{+0.029}_{-0.033}$ \\
\hline

\multicolumn{8}{l}{$\boldsymbol{\mathrm{PAge}+\Omega_{\rm K}}$} \\
DESI+DESY5+SGL             & $69.50^{+2.20}_{-1.90}$ & $0.2650^{+0.0460}_{-0.0380}$ & $\text{---}$ & $0.9487\pm 0.0079$ & $142.8^{+3.7}_{-4.8}$ & $-19.321^{+0.070}_{-0.057}$ & $0.033^{+0.047}_{-0.042}$ \\
DESI+DESY5+CC              & $68.30^{+2.50}_{-2.20}$ & $0.2690\pm 0.0430$ & $\text{---}$ & $0.9498\pm 0.0080$ & $145.2^{+4.1}_{-5.3}$ & $-19.358^{+0.077}_{-0.066}$ & $-0.002\pm 0.050$ \\
DESI+DESY5+GRB             & $73.70\pm 1.30$ & $0.2680\pm 0.0410$ & $\text{---}$ & $0.9501\pm 0.0078$ & $134.7\pm 2.2$ & $-19.194\pm 0.036$ & $0.032\pm 0.044$ \\
DESI+DESY5+SGL+CC          & $69.10^{+1.70}_{-1.50}$ & $0.2670\pm 0.0440$ & $\text{---}$ & $0.9498^{+0.0086}_{-0.0078}$ & $143.6^{+2.9}_{-3.5}$ & $-19.333^{+0.053}_{-0.045}$ & $0.041\pm 0.049$ \\
DESI+DESY5+SGL+CC+GRB      & $72.09\pm 0.98$ & $0.2670\pm 0.0440$ & $\text{---}$ & $0.9524\pm 0.0080$ & $137.8\pm 1.7$ & $-19.243\pm 0.027$ & $0.007\pm 0.038$ \\
DESI+DES-Dovekie+SGL+CC+GRB      & $72.67\pm 0.97$ & $0.3200\pm 0.0400$ & $\text{---}$ & $0.9588\pm 0.0075$ & $137.9\pm 1.7$ & $-19.234\pm 0.027$ & $0.010\pm 0.035$ \\
\hline

\multicolumn{8}{l}{$\boldsymbol{\mathrm{MAPAge}+\Omega_{\rm K}}$} \\
DESI+DESY5+SGL             & $68.60^{+2.20}_{-1.90}$ & $0.0660\pm 0.0960$ & $0.52\pm 0.22$ & $0.9344^{+0.0086}_{-0.0110}$ & $143.9^{+3.8}_{-4.8}$ & $-19.330^{+0.070}_{-0.057}$ & $0.034\pm 0.048$ \\
DESI+DESY5+CC              & $68.20\pm 2.20$ & $0.0830\pm 0.0950$ & $0.49\pm 0.22$ & $0.9362\pm 0.0098$ & $144.6^{+4.1}_{-4.9}$ & $-19.343^{+0.073}_{-0.064}$ & $0.003\pm 0.050$ \\
DESI+DESY5+GRB             & $73.40\pm 1.30$ & $0.0870\pm 0.0920$ & $0.48\pm 0.21$ & $0.9370\pm 0.0098$ & $134.7\pm 2.1$ & $-19.186\pm 0.035$ & $0.029\pm 0.045$ \\
DESI+DESY5+SGL+CC          & $68.60\pm 1.60$ & $0.0670\pm 0.0910$ & $0.53\pm 0.21$ & $0.9349\pm 0.0097$ & $143.9^{+2.9}_{-3.4}$ & $-19.330^{+0.050}_{-0.044}$ & $0.033\pm 0.048$ \\
DESI+DESY5+SGL+CC+GRB      & $71.57\pm 0.97$ & $0.0990\pm 0.0950$ & $0.48\pm 0.21$ & $0.9394\pm 0.0098$ & $138.1\pm 1.6$ & $-19.241\pm 0.025$ & $0.001\pm 0.038$ \\
DESI+DES-Dovekie+SGL+CC+GRB      & $72.23\pm 0.96$ & $0.1810\pm 0.0970$ & $0.37^{+0.23}_{-0.21}$ & $0.9482\pm 0.0098$ & $138.2\pm 1.6$ & $-19.234\pm 0.025$ & $0.005\pm 0.037$ \\
\hline
\end{tabular}
}
\end{table*}

\section{Results and discussions}\label{sec3}

\begin{figure*}[htbp]
\includegraphics[scale=0.65]{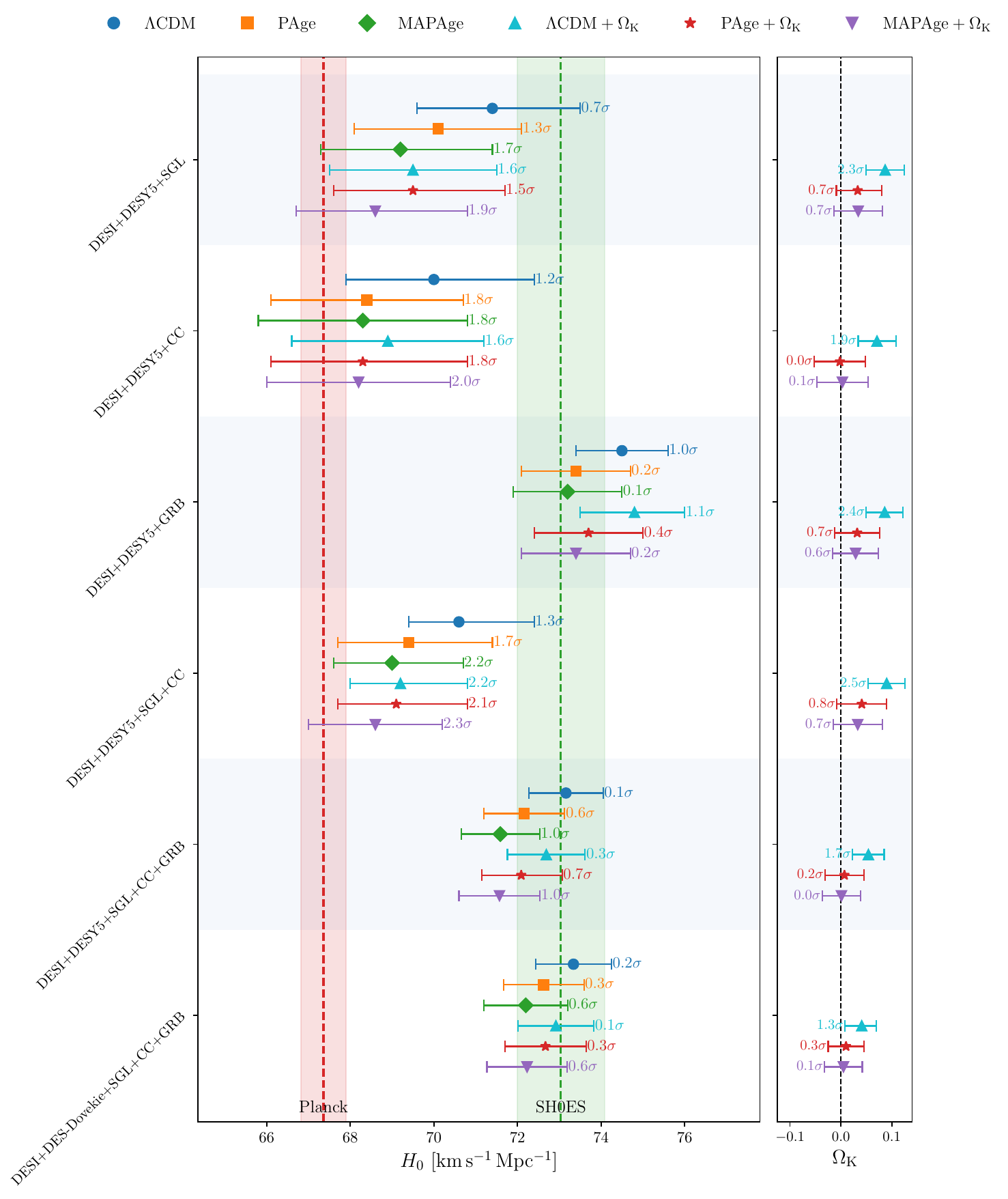}
\centering
\caption{\label{fig1} The whisker plot of $H_0$ in the $\Lambda$CDM, PAge, MAPAge, $\Lambda\mathrm{CDM}+\Omega_{\rm K}$, PAge$+\Omega_{\rm K}$, and MAPAge$+\Omega_{\rm K}$ models using the DESI, DESY5, DES-Dovekie, SGL, CC, and GRB data. Here, for $H_0$ we calculate the tension with SH0ES, while for $\Omega_{\rm K}$ we assess its deviation from zero.}
\end{figure*}

\begin{table*}[htbp]
\renewcommand\arraystretch{1.5}
\centering
\caption{The summary of $\ln\mathcal{B}_{ij}$ using the DESI, DESY5, DES-Dovekie, SGL, CC, and GRB data. A positive value of $\ln\mathcal{B}_{ij}$ indicates a preference for model $i$ over the $\Lambda$CDM + $\Omega_{\rm K}$ model, where $i$ denotes the $\Lambda$CDM, PAge, MAPAge, PAge$+\Omega_{\rm K}$, or MAPAge$+\Omega_{\rm K}$ models.}
\label{table:Bayes_factors}
\resizebox{1.0\textwidth}{!}{
\setlength{\tabcolsep}{8pt}
\begin{tabular}{lcccccc}
\hline\hline
Data & $\Lambda\mathrm{CDM}$ & $\mathrm{PAge}$ & $\mathrm{MAPAge}$ & $\Lambda\mathrm{CDM}+\Omega_{\rm K}$ & $\mathrm{PAge}+\Omega_{\rm K}$ & $\mathrm{MAPAge}+\Omega_{\rm K}$ \\
\hline
DESI+DESY5+SGL             & $-0.99\pm0.50$  & $-0.46\pm0.53$  & $0.50\pm0.48$  & $0$  & $-1.40\pm0.48$  & $-1.97\pm0.48$  \\
DESI+DESY5+CC              & $-0.04\pm0.50$  & $0.74\pm0.49$   & $1.22\pm0.48$  & $0$  & $-1.44\pm0.48$  & $-1.04\pm0.47$  \\
DESI+DESY5+GRB             & $-1.17\pm0.53$  & $-0.40\pm0.59$  & $0.78\pm0.49$  & $0$  & $-1.69\pm0.47$  & $-1.00\pm0.48$  \\
DESI+DESY5+SGL+CC          & $-1.26\pm0.51$  & $-0.67\pm0.51$  & $0.09\pm0.49$  & $0$  & $-0.97\pm0.48$  & $-0.79\pm0.48$  \\
DESI+DESY5+SGL+CC+GRB      & $0.73\pm0.51$   & $2.74\pm0.51$   & $3.64\pm0.49$  & $0$  & $0.24\pm0.50$   & $1.07\pm0.49$   \\
DESI+DES-Dovekie+SGL+CC+GRB & $1.87\pm0.54$  & $2.49\pm0.53$   & $3.30\pm0.52$  & $0$  & $0.73\pm0.52$   & $0.98\pm0.51$  \\
\hline
\end{tabular}
}
\end{table*}

\begin{figure*}[htbp]
\includegraphics[scale=0.7]{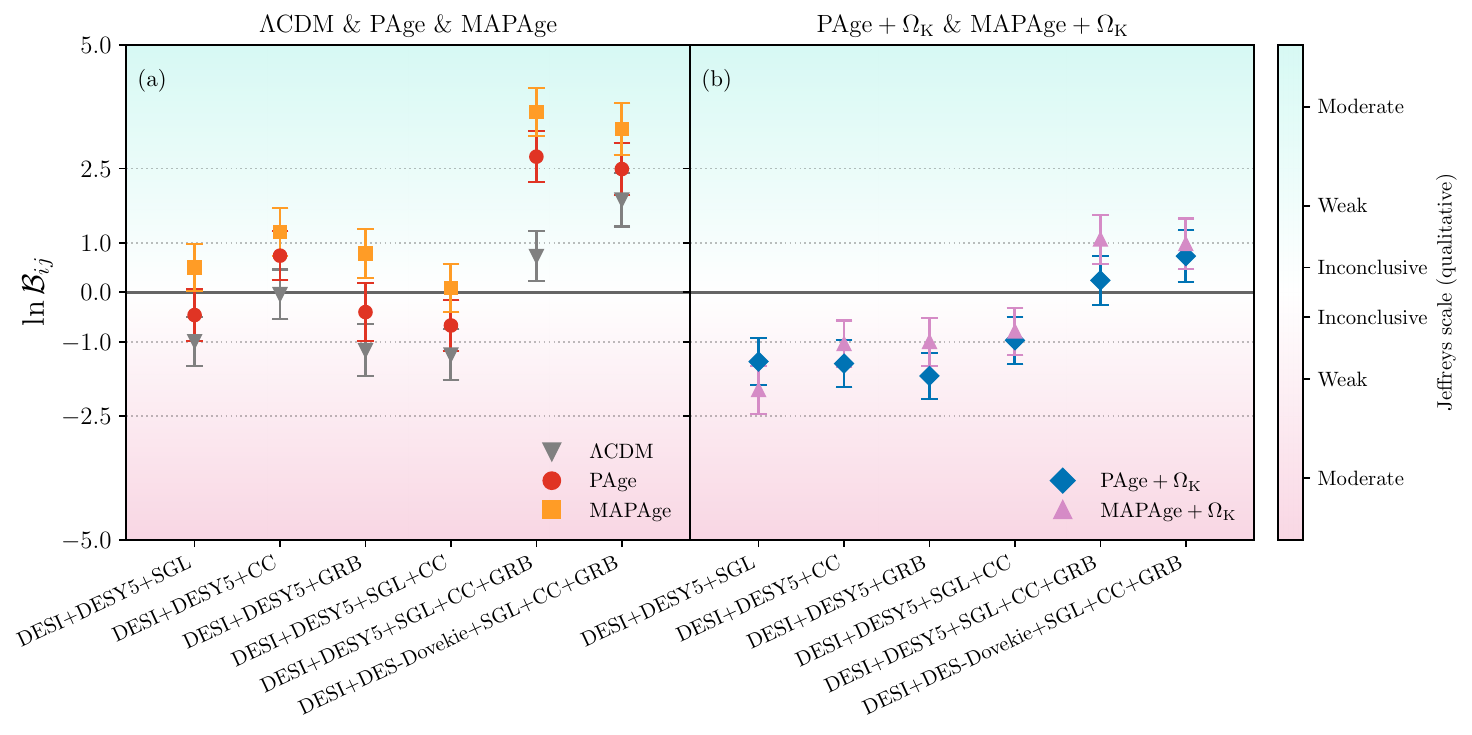}
\centering
\caption{\label{fig2} The logarithmic Bayes factors ($\ln\mathcal{B}_{ij}$) obtained from the DESI, DESY5, DES-Dovekie, SGL, CC, and GRB data, where $i$ denotes the $\Lambda$CDM, PAge, MAPAge, PAge$+\Omega_{\rm K}$, and MAPAge$+\Omega_{\rm K}$ models and $j$ is the $\Lambda$CDM$+\Omega_{\rm K}$ model.}
\end{figure*}

We first focus on the $H_0$ measurements obtained by calibrating DESI+DESY5 individually using SGL, CC, and GRB in the scenario of $\Omega_\mathrm{K}=0$. As shown in Fig.~\ref{fig1}, the PAge and MAPAge models generally yield lower $H_0$ values than $\Lambda$CDM. This behavior can be attributed to the fact that PAge and MAPAge are cosmology independent parameterizations and thus may more faithfully reflect the trend present in the current DESI data, namely a preference for a lower $H_0$ that could arise from possible dark energy dynamics. Moreover, the $H_0$ values inferred from SGL are slightly higher than those from CC and exhibit marginally better precision, and SGL and CC are consistent within $1\sigma$ while both show a $0.7-1.8\sigma$ offset relative to the SH0ES measurement. In contrast, GRB yields larger $H_0$ values which are in tension with SH0ES at the $0.1-1.0\sigma$ level, and even slightly exceeding SH0ES. This discrepancy may stem from intrinsic systematic uncertainties in the GRB measurements, such as selection effects in the GRB sample. Alternatively, it could originate from potential systematics introduced during the calibration of GRBs with low-redshift SNe. The GRB dataset in our analysis was calibrated using the low-redshift Union2 SNe sample \cite{Demianski:2019vzl}. Because this calibration anchors the GRB distance moduli to low-redshift SNe, they naturally inherit the characteristics of the local distance ladder, consequently tending to favor higher $H_0$ values. It is noteworthy that the GRB-based $H_0$ determinations provide the most precise constraints compared to those from SGL and CC, which is likely due to the wide redshift coverage of the GRB sample.

Next, we combine SGL and CC, which provide mutually consistent $H_0$ estimates, to construct an IDL. Using DESI+DESY5+SGL+CC, the precision of the combined $H_0$ measurements improves by roughly $25\%$ relative to using SGL or CC alone. Finally, combining GRB, SGL, and CC yields high-precision late-time measurements of $H_0$ for $\Lambda$CDM, PAge, and MAPAge. Under these combinations, the tension with SH0ES is reduced to approximately $0.1\sigma$, $0.6\sigma$, and $1.0\sigma$ for $\Lambda$CDM, PAge, and MAPAge, respectively, and the resulting measurement precision surpasses that of SH0ES. Notably, the combination of GRB with SGL and CC is justified because, despite some differences among the individual measurements, the single measurement uncertainties are relatively large. By computing pairwise tensions in the measurement of $H_0$, we observe at most $1.7\sigma$ tension between GRB and SGL and at most $1.9\sigma$ between GRB and CC, both below the conventional $3\sigma$ significance threshold. Furthermore, from the measurements of $r_\mathrm{d}$ and $M_\mathrm{B}$, we observe their strong degeneracies with $H_0$ that are naturally present in the IDL. For example, results from DESI+DESY5+SGL+CC+GRB in PAge yield corresponding $H_0$ values broadly consistent with SH0ES. In contrast, results from DESI+DESY5+CC are consistent with the Planck measurement.

When we further consider non-zero curvature in the $\Lambda\mathrm{CDM}+\Omega_\mathrm{K}$ model, the $\Omega_\mathrm{K}$ estimates from various data combinations (incorporating SGL, CC, and GRB alongside DESI+DESY5) show deviations from zero ranging from $1.7\sigma$ to $2.5\sigma$. These deviations arise because the current DESI and DESY5 data already show a significant deviation within the $\Lambda\mathrm{CDM}$ framework. Therefore, treating curvature as a free parameter in $\Lambda\mathrm{CDM}$ reveals an approximately $2\sigma$ level preference for positive curvature. A similar conclusion is also reported by the DESI Collaboration utilizing CMB+DESI data. If non-zero curvature is instead considered under the cosmology-independent PAge and MAPAge approximations, this positive-curvature preference disappears, and all data combinations are consistent with $\Omega_\mathrm{K}=0$ within $1\sigma$. Therefore, the IDL constructed from current late-time data strongly supports a spatially flat universe in a model-independent sense.

Furthermore, noting the recent recalibration of the DES SN data aimed at resolving internal consistency issues between low- and high-redshift samples, we also utilize the new DES-Dovekie sample~\cite{DES:2025sig}. As shown in Table~\ref{table:St_new} and Fig.~\ref{fig1}, updating from DESY5 with DES-Dovekie leads to a slight upward shift in the constrained $H_0$ values across all models; for example, $H_0$ increases from $71.59 \pm 0.94$ to $72.20 \pm 1.00~{\rm km}~{\rm s}^{-1}~{\rm Mpc}^{-1}$ in the MAPAge model. In addition, the spatial curvature $\Omega_{\rm K}$ remains robustly constrained around zero ($\Omega_{\rm K} = 0.005 \pm 0.037$ in the MAPAge$+\Omega_{\rm K}$ model). Overall, updating the SN sample to DES-Dovekie has almost no impact on our main findings.

Finally, we compare model preference using Bayes factors $\ln\mathcal{B}_{ij}$ based on the current data. From Table~\ref{table:Bayes_factors} and the left panel of Fig.~\ref{fig2}, we find that PAge and MAPAge are generally preferred relative to $\Lambda\mathrm{CDM}+\Omega_{\rm K}$ for all data combinations. Specifically, MAPAge is more strongly favored at a level ranging from weak to moderate evidence relative to PAge. For instance, DESI+DES-Dovekie+SGL+CC+GRB gives a $\ln\mathcal{B}_{ij} = 3.30\pm0.52$ for MAPAge relative to $\Lambda\mathrm{CDM}+\Omega_{\rm K}$, indicating moderate evidence favoring MAPAge. When non-zero curvature is included, the preference for PAge and MAPAge decreases for all data combinations, as shown in the right panel of Fig.~\ref{fig2}. For MAPAge$+\Omega_{\rm K}$, DESI+DES-Dovekie+SGL+CC+GRB yields $\ln\mathcal{B}_{ij}=0.98\pm0.51$, indicating a comparable favor to $\Lambda\mathrm{CDM}+\Omega_{\rm K}$.

\section{Conclusion}\label{sec4}
In this letter, we construct a PAge-improved IDL using the latest late-universe data and obtain high-precision $H_0$ and $\Omega_\mathrm{K}$ measurements by adopting the PAge and MAPAge parameterizations. Using DESI+DES-Dovekie+SGL+CC+GRB, we obtain $H_0 = 72.63 \pm 0.96\,\mathrm{km\,s^{-1}\,Mpc^{-1}}$ and $H_0 = 72.20 \pm 1.00\,\mathrm{km\,s^{-1}\,Mpc^{-1}}$ for the PAge and MAPAge models, reducing the $H_0$ tension to $0.3\sigma$ and $0.6\sigma$, respectively. Moreover, allowing non-zero curvature within the $\Lambda$CDM framework yields an approximately $2\sigma$ preference for positive $\Omega_{\rm K}$. However, when employing the model-independent PAge and MAPAge models, this preference disappears, and the constraints become fully consistent with $\Omega_\mathrm{K} = 0$ within $1\sigma$, indicating that current late-universe data support a spatially flat universe. Finally, Bayesian comparisons demonstrate that PAge and MAPAge are favored over the $\Lambda\mathrm{CDM}+\Omega_{\rm K}$ model by the current data.

In summary, current late-universe probes independently provide $H_0$ measurements competitive with CMB-based measurements and consistently support a flat universe with $\Omega_\mathrm{K} = 0$. It is anticipated that the full future DESI dataset, combined with more precise late-universe observations from CSST~\cite{CSST:2025ssq}, Euclid~\cite{Euclid:2024yrr}, and LSST~\cite{LSST:2008ijt}, will provide independent measurements of $H_0$ and $\Omega_\mathrm{K}$ with significantly improved precision.

\section*{Acknowledgments}
We thank Sheng-Han Zhou, Yi-Min Zhang, and Peng-Ju Wu for their helpful discussions. This work was supported by the National Natural Science Foundation of China (Grants Nos. 12533001, 12575049, and 12473001), the National SKA Program of China (Grants Nos. 2022SKA0110200 and 2022SKA0110203), the China Manned Space Program (Grant No. CMS-CSST-2025-A02), and the 111 Project (Grant No. B16009).
\bibliography{main}

\end{document}